\documentclass[12pt,notoc]{JHEP3}

\usepackage{amsmath,amssymb,euscript,array,cite,mathrsfs}

\usepackage{epsfig}

\def\XXint#1#2#3{{\setbox0=\hbox{$#1{#2#3}{\int}$}
     \vcenter{\hbox{$#2#3$}}\kern-.5\wd0}}

\def\AA{\mathscr{A}}
\def\BB{\mathscr{B}}

\newcommand{\IM}{\operatorname{Im}}

\def\c{\gamma}
\def\d{\delta}

\def\l{\lambda}

\def\w{\omega}

\def\D{\Delta}

\def\det{{\rm det}}

\def\Dbarslash{\,\,{\raise.15ex\hbox{/}\mkern-12mu {\bar D}}}
\def\Dslash{\,\,{\raise.15ex\hbox{/}\mkern-12mu D}}
\def\delslash{\,\,{\raise.15ex\hbox{/}\mkern-9mu \partial}}
\def\delbarslash{\,\,{\raise.15ex\hbox{/}\mkern-9mu {\bar\partial}}}

\newcommand{\EQ}[1]{\begin{equation}\begin{split} #1 \end{split}\end{equation}}

\newcommand{\SP}[1]{\begin{equation}\begin{split} #1
\end{split}\end{equation}}

\title{The Effect of Gravitational Tidal Forces on Vacuum
Polarization: How to Undress a Photon}
\author{Timothy J. Hollowood and Graham M. Shore\\
Department of Physics,\\ University of Wales Swansea,\\
Swansea, SA2 8PP, UK.\\
E-mail: {\tt t.hollowood@swansea.ac.uk, g.m.shore@swansea.ac.uk}}
\abstract{
The effect of gravitational tidal forces on photon propagation in 
curved spacetime is investigated. It is found that the imaginary part
of the local refractive index $\IM n(u;\w)$ may be negative as well
as positive, corresponding to a local amplification as well as 
attenuation of the amplitude of the renormalized photon field. 
This is interpreted in terms of the effect of tidal forces on the 
virtual $e^+e^-$ cloud surrounding the bare photon field---a 
positive/negative $\IM n(u;\w)$ corresponds to an increased 
dressing/undressing of the bare photon. Below threshold decays
of the photon to $e^+e^-$ pairs can occur. Photon undressing in the
vicinity of a black hole singularity is described as an example.
These results are shown to be consistent with unitarity and the
optical theorem in curved spacetime, which is derived here both in
a local form and integrated over the photon trajectory.
}

\begin{document}

\noindent{\bf 1. Introduction}

A consequence of vacuum polarization in QED in curved spacetime is 
that photons do not simply propagate along classical null geodesic 
trajectories. Instead, the spacetime acts as a medium with 
a non-trivial refractive index. In a series of papers 
\cite{Hollowood:2007kt,Hollowood:2007ku,Hollowood:2008kq,Hollowood:2009qz}, 
we have developed the theory of photon propagation in QED in curved spacetime
and calculated the full frequency dependence of the refractive 
index in terms of a geometric quantity, the Van Vleck-Morette (VVM)
determinant, which characterizes the null geodesic congruence around 
the classical trajectory. This has resolved the long-standing puzzle 
originating from the discovery of Drummond and Hathrell 
\cite{Drummond:1979pp} that at low frequencies, the phase velocity 
may be superluminal. The geometry induces a novel analytic 
structure for the Green functions and refractive index which modifies the conventional Kramers-Kronig dispersion relation and removes the apparent 
paradox between low-frequency superluminal motion and causality
\cite{Hollowood:2008kq}.

An initially surprising result of this work was that in some backgrounds, 
the imaginary part of the refractive index may be negative. In a 
conventional optical medium, ${\rm Im}~n(\w)$ is always positive, 
corresponding to scattering of photons from the beam and a reduction 
in the amplitude. A negative ${\rm Im}~n(\w)$ implies gain, with energy 
pumped in from an external source. In quantum field theory, ${\rm Im}~n(\w)$ 
is related via the optical theorem to the rate of production of $e^+ e^-$ 
pairs and therefore vanishes; even for a massive ``photon'' above the
pair production threshold, ${\rm Im}~n(\w)$ would be manifestly positive. 
The fact that ${\rm Im}~n(\w)$ can be negative in curved spacetime, 
and can be non-vanishing and positive below the $e^+ e^-$ threshold, 
requires a careful explanation and interpretation. In particular, it 
involves a reappraisal of the optical theorem in curved spacetime. 

In this paper, we summarize the resolution of this problem, leaving
the full formal development to a longer paper \cite{NEW}.
We also provide an intuitive picture of how the amplitude may either
increase or decrease depending on the sign of ${\rm Im}~n(\w)$ based
on the idea of ``dressing'' a renormalized quantum field.

In QED, vacuum polarization means we may regard the renormalized photon
field as the bare field dressed with a cloud of virtual $e^+ e^-$ pairs.
Since the electron has a mass, this cloud has a scale characterized by
the electron Compton wavelength $\l_c = h/mc$, so the renormalized photon 
should be thought of as an extended body with size $\l_c$ propagating through
curved spacetime with scale $L \sim 1/\sqrt{\cal R}$, where ${\cal R}$ 
is a typical curvature. Classically, massless point particles follow null 
geodesics in curved spacetime. Extended bodies, on the other hand, are 
subject to gravitational tidal forces because nearby geodesics can diverge 
or converge. In quantum field theory, because of vacuum polarization,
we therefore expect that when $L$ becomes comparable to $\l_c$, these 
tidal forces will distort the virtual cloud and affect the degree of 
dressing of the renormalized photon.

\begin{figure}[ht] 
\centerline{\includegraphics[width=2in]{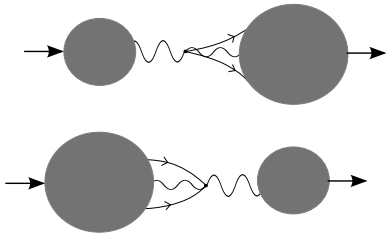}}
\caption{\footnotesize Diagrams which illustrate the 
the cases where the photon amplitude decreases (top) and
increases (bottom) arising from the production of more/less virtual
$e^+ e^-$ pairs, respectively. The top/bottom cases occurs when
the rate of change of the acceleration on nearby geodesics is
positive/negative.}\label{f3}
\end{figure}

The conceptual picture that we shall confirm is the following. 
To linear order in the curvature, when the rate of change of the 
acceleration of nearby geodesics is positive---in the direction of
increased stretching---more virtual $e^+ e^-$ pairs are produced and 
the photon's amplitude decreases. This is illustrated in the top 
diagram of Fig.~\ref{f3} which indicates that the photon is becoming 
``more dressed''. If the rate of change of the acceleration is at least 
constant over a finite region along the geodesic then this can interpreted 
as real pair production and the amplitude reduces at a rate which is at 
least as great as exponential. 
On the contrary, when the rate of change of the acceleration of 
nearby geodesics is negative---in the direction of increased 
squeezing---virtual $e^+ e^-$ pairs are forced back into the photon. 
In this case the amplitude of is amplified. This is illustrated in the 
lower diagram of Fig.~\ref{f3}.

This mechanism gives rise to a perturbative contribution to ${\rm Im}~n(\w)$
of order $\w \partial_u {\cal R}/m^4$ 
(where $u$ is a lightcone coordinate along the null geodesic)
in cases where the curvature is not constant along the photon trajectory.
${\rm Im}~n(\w)$ can be positive or negative and the photon amplitude 
can decrease (even below threshold) or increase. In many respects, 
the situation resembles the transient phase immediately following
an interaction being turned on in conventional flat spacetime QFT.
In both cases, the possibility of amplification as well as attenuation 
originates in the loss of translation invariance, which enables energy 
conservation to be circumvented.

In addition to this very intuitive mechanism there is a further effect,
non-perturbative in $\w\sqrt{\cal R}/m^2$, which can give rise 
to an imaginary part for the refractive index. This is related to the 
occurrence of {\it conjugate points} along the photon's null geodesic $\c$. 
Two points on $\c$ which can be linked by a 
continuous family of geodesics arbitrarily close to $\c$ are said to
be conjugate. The existence of these nearby geodesics implies there are 
zero modes in the fluctuations around the classical path and
consequently singularities in the VVM matrix 
\cite{Hollowood:2007kt,Hollowood:2007ku,Hollowood:2008kq}. 
This is the ultimate geometric origin 
of the novel analytic structure for the refractive index in curved spacetime and 
in some cases may imply the existence of a non-vanishing contribution
to ${\rm Im}~n(\w)$. This effect is present even in the simple example
of symmetric plane waves, or Cahen-Wallach spaces \cite{Cahen}, where
the curvature is constant along the photon path. In these cases, 
${\rm Im}~n(\w)$ is always positive,\footnote{
Note that in refs.\cite{Hollowood:2008kq,Hollowood:2009qz}, there is an overall
sign error in the refractive index for scalar QED. In particular, $\IM n(\w)$
for scalar QED in a Ricci-flat symmetric plane wave should be positive.
} a result which
can now be understood from our analysis of the optical theorem in
curved spacetime. For other backgrounds, there is no restriction on 
the sign of ${\rm Im}~n(\w)$.

Clearly, the geometry of fluctuations around the classical photon trajectory
is central to this analysis.
Consider a null congruence centred on $\c$ and let $z^i(u), i=1,2$ be 
the two spacelike connecting vectors linking neighbouring geodesics.
The $z^i(u)$ satisfy the Jacobi equation 
\EQ{
\ddot z^i+h_{ij}(u)z^j=0\ 
\label{aa}
}
where $h_{ij}(u)=R^i{}_{uju}(u)$, the components of the Riemann tensor
in a frame associated to the Brinkmann coordinates $(u,v,z^i)$ 
adapted to the geodesic $\gamma$, which is the curve $v=z^i=0$. 
The metric in the neighbourhood of $\gamma$ takes the form
\EQ{
ds^2=2du\,dv-h_{ij}(u)z^i z^j\,du^2
+dz^i\,dz^i\ .
\label{ab}
}
This metric is known as the Penrose limit for the given background
and null geodesic $\gamma$ and encodes the geometry of geodesic deviation
\cite{Penrose,Blau2}. 
For later use, we also introduce the Rosen
coordinates $(u,V,x^a)$, $a=1,2$, in which the approximate metric
\eqref{ab} takes the form
\EQ{
ds^2=2du\,dV+C_{ab}(u)dx^a\,dx^b\ .
\label{ac}
}
The affine parameter $u$ along $\gamma$ is common to both sets of coordinates.

A key role is played by the Van Vleck-Morette determinant
$\Delta(x,x')$ evaluated at the two points $x,x'\in\gamma$ with affine
parameters $u$ and $u'$, so we write $\Delta(u,u')$. 
There are different ways to define the VVM determinant but one
very physical way is via solutions of the geodesic equation
\eqref{aa}. If we write the solution for the Jacobi field $z^i(u)$ in terms 
of some initial data at $u'$,
\EQ{
z^i(u) =\BB^i{}_j(u,u') z^j(u') + \AA^i{}_j(u,u')\dot z^j(u') \ ,
\label{ad}
}
then \cite{Hollowood:2008kq,Hollowood:2009qz}
\EQ{
\Delta(u,u')=\frac{(u-u')^2}{\det\AA(u,u')}\ .
\label{ae}
}

\vskip0.5cm
\noindent{\bf 2. Vacuum Polarization and the Refractive Index}
\vskip0.2cm
The Penrose limit will be a good approximation for a general curved spacetime
provided the curvature scale is greater than the electron Compton wavelength,
i.e. $ L \gg \l_c$, or equivalently ${\cal R}/m^2 \ll 1$, and
we work in the geometric optics, or eikonal, approximation
where the curvature scale is large compared to the photon wavelength, i.e.
$L \gg \l$, or ${\cal R}/\w^2 \ll 1$. We refer to these two conditions as the 
``WKB limit''. This leaves the dimensionless combination $\w\sqrt{\cal R}/m^2$ 
of the three scales as a free parameter in our calculation of the full 
frequency dependence of $n(\w)$.

To simplify the presentation and focus on the key issues of the sign of
${\rm Im}~n(\w)$ and the optical theorem, we quote results here for a model
field theory comprising a massless scalar (photon) $A$ interacting with
two massive scalars (electrons) $\phi$ through an $e A\phi^2$ coupling.
The case of QED involves some further subtleties with renormalization
and is described elsewhere \cite{NEW}.

In the leading eikonal approximation, the solution of the free wave equation 
$\square A^{(0)}(x)=0$ takes the approximate form
\EQ{
A^{(0)}(x)=g(x)^{-1/4}\, e^{i\omega \vartheta(x)}\ ,
\label{ba}
}
where $g=\det\, g_{\mu\nu}$ and $k^\mu=\partial^\mu \vartheta(x)$ is the tangent
vector to a null congruence $g(k,k)=0$. Along the geodesics $\vartheta(x) = V$ 
is constant while the affine parameter $u$ varies. If we pick out $\gamma$ to be
the geodesic with $V=0$ then in the neighbourhood of $\gamma$,
$V$ is identified with the Rosen coordinate. 

In our model field theory, vacuum polarization is given by the usual
QED-like one-loop graph of ${\cal O}(e^2)$ with the scalars $\phi$ in the 
loop and external $A$ lines. Since the usual notion of a particle is 
inherently ambiguous in a general curved spacetime, we shall focus on the
behaviour of the field itself and use the techniques of
non-equilibrium QFT to set-up an initial value problem. We therefore
imagine that on the null surface $u=u_0$ the coupling constant
$e$ is turned on instantaneously and then watch as the bare
field evolves along the null direction $u$.\footnote{Strictly
  speaking a null surface is not a Cauchy surface. However, in the
  philosophy of light-cone quantization one can imagine defining
  boundary conditions in the transverse directions $(V,x^a)$ which
  have the effect of giving a well-defined initial value problem.} 
The field satisfies a non-local equation of the form
\EQ{
\square\, A(x) +  e^2\int d^4x'\,
\sqrt{g(x')}\,\Pi_R(x,x')A(x')=0\ ,
\label{bb}
}
where $\Pi_R(x,x')$ is the one-loop retarded self-energy. Since the
latter is only non-vanishing when $x'$ lies in the past light-cone of
$x$, the equation is manifestly causal. 
The aim is to compute the solution of this perturbatively 
in $e$ starting with the classical solution
\eqref{ba} in the form
\EQ{
A(x)=A^{(0)}(x)\big(1+ie^2\,{\cal Q}^{(1)}(x)+\cdots\big)\ .
\label{bc}
}
The one-loop retarded self-energy
is 
\EQ{
\Pi_R(x,x')= i\theta(u-u')\IM\,G_F(x,x')^2\ ,
\label{bd}
}
where $G_F(x,x')$ is the Feynman propagator of 
$\phi(x)$, and in order to extract the one-loop correction we need to
calculate the integral
\EQ{
\int d^4x'\,
\sqrt{g(x')}\,\Pi_R(x,x')A^{(0)}(x')\ .
\label{be}
}
This integral becomes tractable in the WKB limit, where we can replace
a general background spacetime by its Penrose limit \eqref{ab} or \eqref{ac}.

First, we introduce some notation which will be useful when we discuss
the optical theorem. The general on-shell modes for a classical field 
of mass $m$ in a plane wave background in Rosen coordinates are
\EQ{
\phi_{p}(x)=g(u)^{-1/4}\exp\left[i\omega V+ip_ax^a
-\tfrac i{2\omega}(\psi^{ab}(u)p_ap_b + m^2 u)\right]\ ,
\label{bf}
}
where $p = (\w,p_a)$ includes a transverse momentum and we have defined 
the $2\times 2$ matrix
\SP{
\psi(u)=\int^u du'\,C(u')^{-1}\ .
\label{bg}
}
This matrix is related to the VVM determinant via
\EQ{
\Delta(u,u')=\frac1{\sqrt{g(u)g(u')}}\cdot\frac{(u-u')^2}{\det\,\psi(u,u')}\ .
\label{bh}
} 
These modes satisfy the orthonormality property,
\EQ{
(\phi_{p'},\phi_p) \equiv i \int d^3x \sqrt{g}~\phi_{p'}^*(x)
\overleftrightarrow{\partial_V} \phi_p(x) ~=~
2\w (2\pi)^3 \d(\w-\w') \d^{(2)}(p_a - p_a^{\prime}) \ ,
\label{bi}
}
where the integral is over $x=(V,x^a)$.
The original massless photon modes $A^{(0)}$ of \eqref{ba} were chosen 
with $p_a=0$.

Now define the partial ``Fourier transform'' of the self-energy with respect
to these (massless) modes:
\SP{
&(2\pi)^3 \d(\w-\w') \d^{(2)}(p_a - p_a^{\prime})~
\tilde\Pi_R(u,u';\w,p_a) \\
&~~~~~~~~~~~~ =~~\int d^3x\sqrt{g}\,\int d^3x'\sqrt{g'}~
\phi_p^*(x) \Pi_R(x,x') \phi_{p'}(x') \ .
\label{bj}
}
Solving the field equation \eqref{bb}, and taking into account that
$\Pi_R(x,x')$
has support only for $u' < u$,
we can then write the quantum correction ${\cal Q}^{(1)}$ of \eqref{bc}
in the form:
\EQ{
{\cal Q}^{(1)}(u) ~=~ {1\over2\w}\, \int_{u_0}^u du^{\prime\prime}\,
\int_{u_0}^{u^{\prime\prime}} du'\, \tilde\Pi_R(u^{\prime\prime},u';\w,0)\ .
\label{bl}
}

Returning to \eqref{be}, we now summarize some of the key points
of the calculation, while the full details can be found in \cite{NEW}:

\vskip0.1cm
\noindent(a)~In the WKB limit, for a fixed $x$ on the null geodesic
$\gamma$, the integral over $x'$ is dominated by a saddle point with $x'\in\gamma$
in the causal past of $x$. To leading order, the metric can be
replaced by the approximate plane wave metric \eqref{ab} or
\eqref{ac} which captures the fluctuations around $\gamma$ to
Gaussian order.

\vskip0.1cm
\noindent(b)~The propagator for a massive scalar field 
in a plane wave is known exactly 
\cite{Gibbons:1975jb}:
\EQ{
G_F(x,x')=i\sqrt{\Delta(x,x')}\int_0^\infty\frac{dT}{(4\pi
  T)^{2}}e^{-im^2T+\tfrac{i\sigma(x,x')}{2T}}\ .
\label{bm}
}
where $\sigma(x,x')$ is the geodesic interval between $x$ and $x'$
and the saddle point over the $x'$, $T_1$ and $T_2$ (one for each
propagator) integrals is at 
\EQ{
V'=x^{\prime a}=0\ ,~~~~u'=u-2m^2\omega\frac{T_1T_2}{T_1+T_2}\ .
\label{bn}
}
What remains is an integral over the Feynman parameter
$0\leq\xi=\frac{T_1}{T_1+T_2}\leq1$ and the point $x'$ on $\gamma$
labelled by affine parameter $u'$ with $u_0\leq u'\leq u$. Performing the
Gaussian integrals around the saddle-point gives the equation for the 
quantum correction ${\cal Q}^{(1)}(x)$. Evaluating this along $\gamma$
gives 
\EQ{
\partial_u{\cal Q}^{(1)}(u)
= \frac1{32\pi^2\omega}
\int_{u_0}^{u}du'\,\frac{\sqrt{\Delta(u,u')}}{u-u'}
\int_0^1d\xi\,e^{-\frac{im^2(u-u')}
{2\omega\xi(1-\xi)}}\ ,
\label{bo}
}
which can easily be integrated with initial condition ${\cal Q}^{(1)}(u_0)=0$
giving
\EQ{
{\cal Q}^{(1)}(u) = {1\over32\pi^2 \w} 
\int_{u_0}^u du^{\prime\prime}\, \int_{u_0}^{u^{\prime\prime}} du' ~ 
\frac{\sqrt{\Delta(u^{\prime\prime},u')}}{u^{\prime\prime}-u'}
\int_0^1d\xi\,e^{-\frac{im^2(u^{\prime\prime}-u')} {2\omega\xi(1-\xi)}}\ .
\label{boo}
}
The curvature is encoded in the Van Vleck-Morette 
determinant. The result \eqref{bo} is divergent for two entirely
different reasons: (i) there is the standard UV divergence as $u'\to
u$. This is the same divergence as in flat space that is absorbed by a
mass counterterm for the $A$ field. This divergence does not affect
the the imaginary part of ${\cal Q}^{(1)}$ which is our main focus
here, so we will not explain this point any further; (ii) 
there is a divergence whenever $u'$ and
$u$ are conjugate points coming from the VVM determinant. 
These divergences are treated by implementing the real space
Feynman prescription, $u-u'\to u-u'-i0^+$. 

Also note that because of the symmetry $\D(x,x') = \D(x',x)$, the
imaginary part of ${\cal Q}^{(1)}(u)$ is given simply by extending the
range of the integration from $u_0 \le u'\le u^{\prime\prime}$
to $u_0 \le u' \le u$. We find
\EQ{
\IM {\cal Q}^{(1)}(u) ~=~
-{i\over64\pi^2 \w}  
\int_{u_0}^u du^{\prime\prime}\, \int_{u_0}^u du' ~ 
\frac{\sqrt{\Delta(u^{\prime\prime},u')}}{u^{\prime\prime}-u'}
\int_0^1d\xi\,e^{-\frac{im^2(u^{\prime\prime}-u')} {2\omega\xi(1-\xi)}}\ .
\label{booo}
}
This quantity appears in the optical theorem, as described in section 3.

\vskip0.1cm
\noindent(c)~In general the one-loop correction can grow as a function of $u$
and it becomes necessary to perform a re-summation using the dynamical
renormalization group \cite{Boyanovsky:2003ui}. 
This results in an exponentiation of the result and
the solution to the quantum corrected field equation \eqref{bb} is then
\EQ{
A(u)=g(u)^{-1/4} e^{i\w V} e^{i e^2 {\cal Q}^{(1)}(u)}
\label{bp}
} 
The result can be interpreted in terms of a position-dependent refractive index 
\EQ{
n(u;\w)=1+\frac {e^2}\omega\partial_u{\cal Q}^{(1)}(u)
\label{bq}
}
with a real part that describes a non-trivial correction to the phase
velocity of the massless field $A(x)$ and an imaginary part that
describes a non-trivial correction to the amplitude
\EQ{
|A(u)|=|A(u_0)|\exp\left[-\omega\int_{u_0}^u du\,\IM n(u;\w)\right]
\label{br}
}
along $\gamma$. 

\vskip0.2cm
The result \eqref{bo} is already non-trivial even in flat space. This
describes a transient region where the bare field
$A$ at $u=u_0$ renormalizes and becomes a dressed field 
in real time. Indeed, as $u$ increases, $\IM \partial_u{\cal Q}^{(1)}(u)$ 
settles down after some oscillations to a positive constant and the ratio
$|A(u)/A(u_0)|$ approaches the wavefunction renormalization factor that
one would calculate in the equilibrium theory:
\EQ{
\left|\frac{A(u)}{A(u_0)}\right| \rightarrow \exp\left[-\frac{e^2}
{6\pi^2m^2}\right]=1-\frac{e^2}
{6\pi^2m^2}+\cdots\ .
\label{bs}
}
Notice that in $A\phi^2$ theory in $d=4$, 
wavefunction renormalization is a finite effect. 
The decrease in the renormalized amplitude can be interpreted in terms
of the intuitive picture of the bare field surrounded by a virtual
cloud of $\phi$ pairs. As this dressing increases, the amplitude falls. 

We can decouple this transient initial dressing of the field from
the effective of gravitational tidal forces by taking $u_0\to-\infty$
and then turning on the curvature at some finite $u$ after the
transients have died out.  However, we must 
not forget the initial dressing process which is essential in
maintaining consistency with unitarity, as explained later in our 
discussion of the optical theorem. 

In the weak curvature limit ${\cal R} \ll m$, we
can expand the VVM determinant to linear order in the curvature
\EQ{
\Delta(u,u')=1+\frac16R_{uu}(u)(u-u')^2-\frac1{12}
\partial_u R_{uu}(u)(u-u')^3+\cdots
\label{bt}
}
where the Ricci tensor $R_{uu}=R^i{}_{uiu}$. Substituting this expansion into the
integral \eqref{bo}, and after UV regularization, 
we determine the correction to the refractive index
\EQ{
n(u;\w)=1-\frac{e^2}{180\pi^2m^4}R_{uu}(u)-
\frac{i e^2\omega}{420\pi^2m^6}\partial_u R_{uu}(u)\ ,
\label{bu}
}
In this approximation, the quantum correction to the amplitude relative
to a point $u'\gg u_0$ is
\EQ{
|A(u)|=|A(u')|\exp\left[\frac{e^2\omega^2}{420\pi^2 m^6}
\big(R_{uu}(u)-R_{uu}(u')\big)\right]\ .
\label{bv}
}
If the Ricci curvature component $R_{uu}(u)$ decreases along $\gamma$,
then from the Jacobi equation \eqref{aa} we see that the rate of
acceleration of the nearby geodesics is positive,  {\it i.e\/}~the
tidal forces are increasing in the direction of stretching the
virtual cloud. In this case,  the amplitude decreases because more
virtual pairs are being produced: $A$ is becoming more dressed.
On the contrary, if $R_{uu}(u)$
increases along $\gamma$, {\it i.e\/}~the tidal forces are increasing in
the direction of squeezing, then the amplitude increases because 
virtual pairs are interacting to produce more $A$: 
$A$ is being ``undressed'' and is returning to its bare state.
These effects are illustrated heuristically in Fig.~\ref{f3}. 

\vskip0.2cm
We now come to the second, non-perturbative, mechanism which can
produce an imaginary part for the refractive index.
If $R_{uu}(u)$ is constant, in which case the approximate metric
\eqref{ab} and \eqref{ac} is a symmetric plane wave, or 
Cahen-Wallach space \cite{Cahen}, then to linear order in the
curvature the amplitude is constant. In fact, $\IM n(\w)=0$ 
to all orders in the curvature expansion. However, in
the case where at least one of the eigenvalues of the 
constant $h_{ij}$ is negative there is a positive contribution to 
$\IM n(\w)$, so a decaying amplitude, which is
non-perturbative in the curvature. The simplest case to consider is
$h_{ij}=-\sigma^2\delta_{ij}$. In this case, the VVM determinant is
\EQ{
\Delta(u,u')=\left[\frac{\sigma(u-u')}{\sinh\sigma(u-u')}\right]^2\ .
\label{bw}
} 
Introducing $t=u-u'$ we have
\EQ{
n(\w)=1+\frac{e^2}{32\pi^2\omega^2}
\int_{0}^{\infty}\frac{dt}{\sinh(\sigma t)}
\int_0^1d\xi\,e^{-\frac{im^2t}
{2\omega\xi(1-\xi)}}\ .
\label{bx}
}
$\IM n(\w)$ can be computed by rotating the
$t=u-u'$ integral $t\to-it+i0^+$ and picking up contributions from the
poles at $t=\frac{n\pi}\sigma$, $n=1,2,\ldots$. This gives
\EQ{
\IM n(\w)=\frac{e^2}{16\pi\omega^2}\int_0^1d\xi\,\frac1{
1+e^{\frac{\pi m^2}{2\omega\sigma\xi(1-\xi)}}}\ ,
\label{by}
}
which is, indeed, non-perturbative in the curvature. In this case the
result is independent of $u$ and so it implies a constant rate of
production of $\phi$ pairs. Such a process is kinematically
disallowed in flat space, but there is no such threshold constraint in
curved space.

This result leads to a very non-trivial check of the formalism in the
following way \cite{NEW}. If we consider the space $dS_3\times R$ with metric
\EQ{
ds^2=-dt^2+\cosh^2(\alpha t)\,\big(d\theta^2+\sin^2\theta\,d\phi^2\big)+d\chi^2\ ,
\label{bz}
}
then the plane wave limit about the geodesic $t=\chi$ and a fixed
point on the $S^2$, {\it e.g.\/}~$\theta=\frac\pi2$, $\phi=0$,
is obtained by transforming
\EQ{
t=\tfrac1{\sqrt2}(u-v)\ ,\qquad \chi=\tfrac1{\sqrt2}(u+v)\ ,
\label{baa}
}
and then expanding in powers of $v$ and $x^1=\theta-\frac\pi2$, $x^2=\phi$:
\EQ{
ds^2\rightarrow 2du\,dv+\cosh^2\big(\tfrac{\alpha u}{\sqrt2}\big)dx^a\,dx^a\ .
\label{bab}
}
This is a symmetric plane wave in Rosen coordinates \eqref{ac} 
with $h_{ij}=-\frac{\alpha^2}{2}\delta_{ij}$.
The non-trivial check on \eqref{by} comes about because the decays
of scalar fields $A\to\phi\phi$ can be calculated exactly in de Sitter
space due to large amount of symmetry as a tree-level process
for any masses. In our case $A$ is massless and using the results in
\cite{Bros:2008sq,Bros:2009bz} 
and taking the limits $\omega\gg\alpha$ and $m\gg\alpha$ one
gets a decay rate which is precisely $2\omega\IM n(\w)$ 
as found in \eqref{by}.

\vskip0.2cm
We therefore see that in curved spacetime, the effect of gravitational
tidal forces on the virtual cloud surrounding the bare field means
that the photon can become increasingly ``undressed'' as well as
dressed as it propagates through spacetime. An especially interesting 
example of this gravitational undressing phenomenon is
photon propagation near a singularity, for example a black hole.
The Penrose limit near the singularity has the universal form 
of a homogeneous plane wave \cite{Blau2}
\EQ{
h_{ij}(u)=\frac{1-\alpha_i^2}{4u^2}\delta_{ij}\ ,
\label{bac}
}
with $\alpha_i$ being either real or imaginary (and choosing $\alpha_i\geq0$ in
the former). For example, the near-singularity Penrose limit of a null
geodesic with non-vanishing angular momentum
in the Schwarzschild solution is
described by such a plane wave with $\alpha_1=\tfrac15$ and
$\alpha_2=\tfrac75$. As the singularity
is approached $u\to0$ (with $u<0$) the $u'$ integral is dominated by
its endpoint $u'=u_0=-\infty$ and we find
\EQ{
\IM {\cal Q}^{(1)}(u)\longrightarrow -|u|^{\frac{\alpha_1+\alpha_2+2}4}\ .
\label{bad}
}
The amplitude therefore increases and the photon becomes progressively
undressed as the singularity is approached. Notice  that the
amplitude remains finite even though $\IM n(u;\w)$ itself diverges if 
$\alpha_1+\alpha_2<2$ as in the case of
Schwarzschild. When either of the $\alpha_i$ is imaginary the
amplitude has a oscillating behaviour.

\vskip0.5cm
\noindent{\bf 3.~~Optical Theorem}
\vskip0.2cm
The final issue we need to address is how the optical theorem
and unitarity can be satisfied if the amplitude can increase
and ${\rm Im}~n(u,\w)$ can be negative. The
answer to this puzzle is that we have only shown that 
the amplitude can increase locally in a region well away from the
initial-value surface. In order to 
verify that unitarity is satisfied, and that the decrease in the
amplitude is related to the probability of the production of $\phi$ pairs,
we have to consider the amplitude over the whole region from the
initial-value surface $u_0$ to $u$. It turns out that while the
amplitude can increase locally, the overall amplitude as measured
from $u_0$ always decreases. The
initial transient dressing of the $A$ field plays an important r\^ole
since it is only a field that is already dressed that can increase its
amplitude by becoming more undressed; a bare field, however, cannot be
undressed any further!

To derive the optical theorem, note first that
\EQ{
\int_{u_0}^u du^{\prime\prime}\IM n(u^{\prime\prime};\w) ~=~
{e^2\over2\w^2} 
\int_{u_0}^u du^{\prime\prime} \int_{u_0}^{u^{\prime\prime}} du'~
\IM \tilde\Pi_R(u^{\prime\prime},u';\w,0) ~=~
{e^2\over\w} \IM {\cal Q}^{(1)}(u) \ .
\label{ca}
}
where $\IM {\cal Q}^{(1)}(u)$ is given in eq.\eqref{booo}.
Now consider the the transition probability for the tree-level
process $A\to\phi\phi$:
\SP{
P_{A\to\phi\phi}(u)~=~
\frac{e^2}{\cal N}\int\frac{d^3p_1}{(2\pi)^3 2\omega_1}\frac{d^3p_2}
{(2\pi)^3 2\omega_2}\left|\int_{u_0}^u
  d^4x\,\sqrt{g}\,A^{(0)}(x)^\dagger\phi_{p_1}(x)\phi_{p_2}(x)\right|^2\ .
\label{cc}
}
Here, $\phi_p(x)$ are the massive on-shell modes defined in \eqref{bf}
and the integrals are over the three-dimensional $p = (\w,p_a)$. 
The normalization factor ${\cal N} = (A^{(0)},A^{(0)}) = 
2\w \d^{(3)}(0)$ cancels against an overall integration on $(V,x^a)$. 
Recall that the Wightman functions 
$G_+(x,x') = \langle 0|\phi(x) \phi(x')|0\rangle$ and
$G_-(x,x') = \langle 0|\phi(x') \phi(x)|0\rangle$ are 
given in terms of these modes by
\EQ{
G_{\pm}(x,x') ~=~ \pm \int_{\omega\gtrless0}\frac{d^3p}{(2\pi)^3 2\omega}~
\phi_p(x) \phi_p(x')^\dagger \ .
\label{cd}
}
and that the frequencies are all positive since they refer to physical
particles in \eqref{cc}.
The retarded self-energy \eqref{bd} is expressed in terms of the
Wightman functions by 
\EQ{
\Pi_R(x,x') ~=~ i \theta(u-u')\bigl[G_+(x,x')^2 
- G_-(x,x')^2\bigr] \ .
\label{ce}
}
A short calculation using the properties $G_+(x,x')^* = G_-(x,x')$
and $G_+(x',x) = G_-(x,x')$, and noting that $\int dV' G_-(x,x')^2 A^{(0)}(x')= 0$
for positive $\w$, now shows that
\EQ{
P_{A\to\phi\phi}(u) ~=~  {e^2\over\w} \int_{u_0}^u du^{\prime\prime}\, 
\int_{u_0}^{u^{\prime\prime}} du' ~ 
\IM \tilde \Pi_R(u^{\prime\prime},u';\w,0) \ ,
\label{cf}
}
and so
\EQ{
P_{A\to\phi\phi}(u) ~=~ 2\w \int_{u_0}^u du^{\prime\prime}
\IM n(u^{\prime\prime};\w) \ .
\label{cg}
}
This is the statement of the optical theorem in curved spacetime.
Unitarity is respected and the integral over the whole trajectory
of $\IM n(u;\w)$ is positive. This shows how unitarity prevents a local amplification of the amplitude from becoming unbounded.

We can also write a local version, defining $\Gamma(u) = \partial_u
P_{A\to\phi\phi}(u)$ as the ``instantaneous rate'' of pair production.
Then,
\EQ{
\Gamma(u) ~=~ 2\w \IM n(u;\w) ~=~ {e^2\over\w} \int_{u_0}^u du'~
\IM \tilde\Pi_R(u,u';\w,0) \ .
\label{ch}
}
Unlike the integrated form, there is no positivity constraint on \eqref{ch}
in general. This explains why there is no conflict with unitarity in the 
examples where we have found $\IM n(u;\w)<0$. However, in cases where 
we have translation invariance along the photon trajectory, as in 
flat spacetime, $n(\w)$ is $u$-independent---as long as one is beyond
the transient region---and $P_{A\to\phi\phi}(u)$ 
is proportional to $u$, so unitarity implies $\IM n(\w)>0$ and
$\Gamma>0$ can be properly interpreted as the rate of 
$A\rightarrow\phi\phi$.

\vskip0.5cm
\noindent{\bf 4.~~Conclusions}
\vskip0.2cm
In conclusion, we have shown how the effect of gravitational tidal
forces on vacuum polarization can alter the dressing of a photon 
as it propagates through space. In particular, we have seen how 
the imaginary part $\IM n(u;\w)$ of the position-dependent refractive 
index can be negative as well as positive, corresponding to 
``undressing'' of the photon rather than the conventional dressing.
Two mechanisms were identified. The first, of order 
$\w\partial_u{\cal R}/m^4$, admits an intuitive interpretation in
terms of curvature variations along the photon's trajectory
altering the balance of the bare field with its virtual $e^+e^-$
cloud, with increasing stretching (squeezing) giving rise to more
dressing (undressing).  The second, which is non-perturbative in
$\w{\cal R}/m^2$, can occur even when the curvature is constant and
is related to the existence of conjugate points on the photon's
null geodesic.

Nevertheless, unitarity is still respected and the optical theorem 
still holds in curved spacetime. In its integrated form \eqref{cf},
\eqref{cg}, it relates the total probability for $e^+e^-$ pair
production to the integral of $\IM n(u;\w)$ along the photon
trajectory, which is manifestly positive.
Except in the special case of translation invariance
along the null geodesic, the corresponding local form \eqref{ch}
has no positivity constraint and the usual interpretation of $\IM n(u;\w)$
as the rate of pair production can break down. However, it does
describe the variation of the amplitude and its interpretation in
terms of dressing and undressing of the photon field. In a sense,
photon propagation in curved spacetime resembles the initial
transient phase in flat spacetime, with the characteristic features
of an oscillating amplitude and below-threshold decay.

In this paper, we have described our results in terms of a scalar
$A\phi^2$ model of QED. In QED itself, there are two further
complications---the photon has a polarization, whose direction influences
the dynamical dressing, and there are subtleties related to
renormalization, requiring a reformulation of the initial-value problem
to circumvent the short-distance divergences at the initial value
surface caused by the lack of asymptotic freedom of QED.
These issues are explored in a more comprehensive treatment of
the effect of gravitational tidal forces on renormalized quantum
fields in ref.\cite{NEW}.


\begin{thebibliography}{99}

{\small

\bibitem{Hollowood:2007kt}
  T.~J.~Hollowood and G.~M.~Shore,
  Phys.\ Lett.\  B {\bf 655} (2007) 67
  [arXiv:0707.2302 [hep-th]].

\bibitem{Hollowood:2007ku}
  T.~J.~Hollowood and G.~M.~Shore,
  Nucl.\ Phys.\  B {\bf 795} (2008) 138
  [arXiv:0707.2303 [hep-th]].

\bibitem{Hollowood:2008kq}
  T.~J.~Hollowood and G.~M.~Shore,
  JHEP {\bf 0812} (2008) 091
  [arXiv:0806.1019 [hep-th]].

\bibitem{Hollowood:2009qz}
  T.~J.~Hollowood, G.~M.~Shore and R.~J.~Stanley,
  JHEP {\bf 0908} (2009) 089
  [arXiv:0905.0771 [hep-th]].

\bibitem{Drummond:1979pp}
  I.~T.~Drummond and S.~J.~Hathrell,
  Phys.\ Rev.\ D {\bf 22} (1980) 343.

\bibitem{NEW}
  T.~J.~Hollowood, G.~M.~Shore, ``{\it The Effect of Gravitational Tidal 
  Forces on Renormalized Quantum Fields\/}'', {to appear\/}.

\bibitem{Cahen}
M.~Cahen and N.~Wallach, ``{\it Lorentzian symmetric spaces\/}'',
Bull. Am. Math. Soc. {\bf 76} (1970) 585-591. 

\bibitem{Penrose}
  R.~Penrose, ``{\it Any space-time has a plane wave as a limit}'', in: 
  Differential geometry and relativity, Reidel and Dordrecht
  (1976), 271-275.

\bibitem{Blau2}
  M.~Blau, ``{\it Plane waves and Penrose limits}'', Lectures given at the
  2004 Saalburg/Wolfersdorf Summer School, 
  {\tt http://www.unine.ch/phys/string/Lecturenotes.html}

\bibitem{Gibbons:1975jb}
  G.~W.~Gibbons,
  Commun.\ Math.\ Phys.\  {\bf 45} (1975) 191.

\bibitem{Boyanovsky:2003ui}
  D.~Boyanovsky and H.~J.~de Vega,
  Annals Phys.\  {\bf 307} (2003) 335
  [arXiv:hep-ph/0302055].

\bibitem{Bros:2008sq}
  J.~Bros, H.~Epstein and U.~Moschella,
  arXiv:0812.3513 [hep-th].

\bibitem{Bros:2009bz}
  J.~Bros, H.~Epstein, M.~Gaudin, U.~Moschella and V.~Pasquier,
  arXiv:0901.4223 [hep-th].

}

\end{thebibliography}
\end{document}